\documentclass[11pt]{article}
\usepackage[a4paper]{geometry}
\usepackage{amsfonts, amsmath, amssymb, amsthm, graphicx, caption, authblk, multirow, makecell, framed, float, xcolor, enumitem, tikz, hyperref}
\theoremstyle{definition}

\theoremstyle{remark}

\newcommand*{\crd}[1]{%
  \framebox{\raisebox{0cm}[0.5\baselineskip][0.05\baselineskip]{%
    \hbox to 0.10cm {\hss#1\hss}}}\hspace{0.05cm}}
   
\newlength{\ralen}
\newcommand{\crdd}[1]{\raisebox{\ralen}{\framebox(9,11){#1}}}

\newcommand{\blk}{\crd{$\alpha$}}
\newcommand{\red}{\crd{$\beta$}}
\newcommand{\back}{\crd{?}}
\newcommand{\backk}{\crdd{?}}
\newcommand{\commitment}[1]{\underbrace{\back\, \back}_{#1}}
\newcommand{\acard}[1]{\overset{#1}{\back}}

\newcommand{\cardbb}[2]{\overset{\crd{#1}}{\crd{#2}}}
\newcommand{\cardc}[3]{\underset{\crd{#2}\, \crd{#3}}{\crd{#1}}}
\newcommand{\cardcc}[3]{\overset{\crd{#1}}{\crd{#2}\, \crd{#3}}}

\newcommand{\backbb}{\cardbb{?}{?}}
\newcommand{\backc}{\cardc{?}{?}{?}}
\newcommand{\backcc}{\cardcc{?}{?}{?}}
\newcommand{\vala}{\cardc{?}{$\alpha$}{$\beta$}}
\newcommand{\valb}{\cardc{?}{$\beta$}{$\alpha$}}
\newcommand{\valc}{\cardc{$\theta_1$}{?}{?}}
\newcommand{\vald}{\cardc{$\theta_2$}{?}{?}}
\newcommand{\vale}{\cardc{$\theta_3$}{?}{?}}
\newcommand{\valf}{\cardc{$\theta_4$}{?}{?}}

\newcommand{\valh}{\cardc{$\theta_6$}{?}{?}}

\begin{document}
\title{Simulating Virtual Players for UNO without Computers}
\author[1]{Suthee Ruangwises\thanks{\texttt{suthee@cp.eng.chula.ac.th}}}
\author[2,3]{Kazumasa Shinagawa\thanks{\texttt{kazumasa.shinagawa.np92@vc.ibaraki.ac.jp}}}
\affil[1]{Department of Computer Engineering, Faculty of Engineering, Chulalongkorn University, Bangkok, Thailand}
\affil[2]{Ibaraki University, Hitachi, Ibaraki, Japan}
\affil[3]{National Institute of Advanced Industrial Science and Technology, Tokyo, Japan}
\date{}
\maketitle

\begin{abstract}
UNO is a popular multiplayer card game. In each turn, a player has to play a card in their hand having the same number or color as the most recently played card. When having few people, adding virtual players to play the game can easily be done in UNO video games. However, this is a challenging task for physical UNO without computers. In this paper, we propose an unconventional protocol that can simulate virtual players using nothing but physical UNO cards. In particular, our protocol can uniformly select a valid card to play from each virtual player's hand at random, or report that none exists, without revealing the rest of its hand. The protocol can also be applied to simulate virtual players in other turn-based card or tile games where each player has to select a valid card or tile to play in each turn.

\end{abstract}

\section{Introduction}
\textit{UNO} is a multiplayer card game invented by Merle Robbins in 1972, and is now commercially produced by a game company Mattel. It is one of the world's most popular card games.

UNO consists of a deck of 108 cards. Except for a few special cards, an UNO card has one of the four colors: red, yellow, green, and blue, and has a single-digit number between 0 and 9 written on the front side. All players have several cards in their hands and take turn to \textit{play} a card, discarding it from the hand face-up.

In each turn, a player plays a card with the same number or color as the most recently played card (with the exception of some special cards). If the player does not have a valid card or does not wish to play, they have to draw a card from the deck and may play that card if it is valid. The first player who plays all cards in their hand wins the game.

UNO can be played by as few as two players, but it is typically played by a large group of players. When having few people, a possible way to make the game more fun is to add virtual players to the game. Most video game versions of UNO provide an option to add bots to the game. However, simulating virtual players is a challenging task for physical UNO without computers.

We consider the most basic model of a virtual player: a randomized one. In each virtual player's turn, it should always play a card if possible. It plays a random card uniformly chosen from all valid cards in its hand, without revealing the rest of the hand. If there is no valid card to play, it should correctly report that without revealing any card in the hand. The main difficulty is how to operate virtual players so that everyone can verify that their moves are correct, while always keeping their hands secret to all players.

\subsection{Related Work}
\textit{Card-based cryptography} is a research area studying unconventional cryptographic protocols that use a deck of physical cards. There are two main lines of research in this area.

\textbf{Secure Multi-Party Computation:} This area involves protocols that can securely compute functions with private input from multiple parties, without revealing the inputs to other parties. Card-based protocols to compute various Boolean functions, including a logical AND function \cite{mizuki16,mizuki09}, a logical XOR function \cite{mizuki09}, a \textit{majority function} \cite{nishida13,toyoda}, and an \textit{equality function} \cite{ruangwises21}, have been developed. Nishida et al. \cite{nishida15} proved that any $n$-variable Boolean function can be computed using $2n+6$ cards. Shinagawa and Nuida \cite{garbled} proved that any Boolean function can be computed using one shuffle. 

\textbf{Zero-Knowledge Proof:} A zero-knowledge proof is an interactive protocol between a prover and a verifier, which allows the prover to show that they know a solution of a specific problem without revealing the solution itself \cite{zkp}. Card-based zero-knowledge proof protocols for a wide range of problems have been developed, including computational problems such as graph isomorphism \cite{graph} and pancake sorting \cite{pancake}, pencil puzzles such as Sudoku \cite{ono,sudoku,tanaka} and Nonogram \cite{nonogram}, and mobile games such as Ball Sort Puzzle \cite{ball}.

Very recently, Shinagawa et al. \cite{oldmaid} developed a card-based player simulation protocol for a card game Old Maid, which can simulate virtual players to play the game with real human players. In particular, their protocol can remove pairs of cards having the same number from each virtual player's hand without revealing the rest of its hand. This result created a new possible line of research: simulating virtual players in card games.

\subsection{Our Contribution}
In this paper, we propose a card-based protocol that can simulate virtual players to play UNO. In particular, our protocol can uniformly select a valid card (according to UNO's rules) to play from each virtual player's hand at random, or report that none exists, without revealing the rest of its hand. Our protocol is the second card-based player simulation protocol after the Old Maid protocol of Shinagawa et al. \cite{oldmaid}.

While Old Maid is purely based on luck, UNO is a strategic game. Knowing hands of virtual players will greaty affect the other players' strategies. Therefore, keeping the hands of virtual players secret is much more crucial in UNO than in Old Maid, highlighting the importance of security of our protocol.

Our protocol is a generic one. We show that it can also be applied to simulate virtual players in other turn-based card or tile games such as \textit{Sevens}, \textit{Hearts}, and \textit{Dominoes}, where each player has to select a valid card or tile to play in each turn.

\section{Preliminaries}
\subsection{Cards}
There are 54 different types of UNO card. 52 of them have color red, yellow, green, or blue on the front side and are called \textit{color cards}. Two of them are black on the front side and are called \textit{black cards}. All cards have indistinguishable back sides.

\begin{figure}[H]
\centering
\includegraphics[width=80mm]{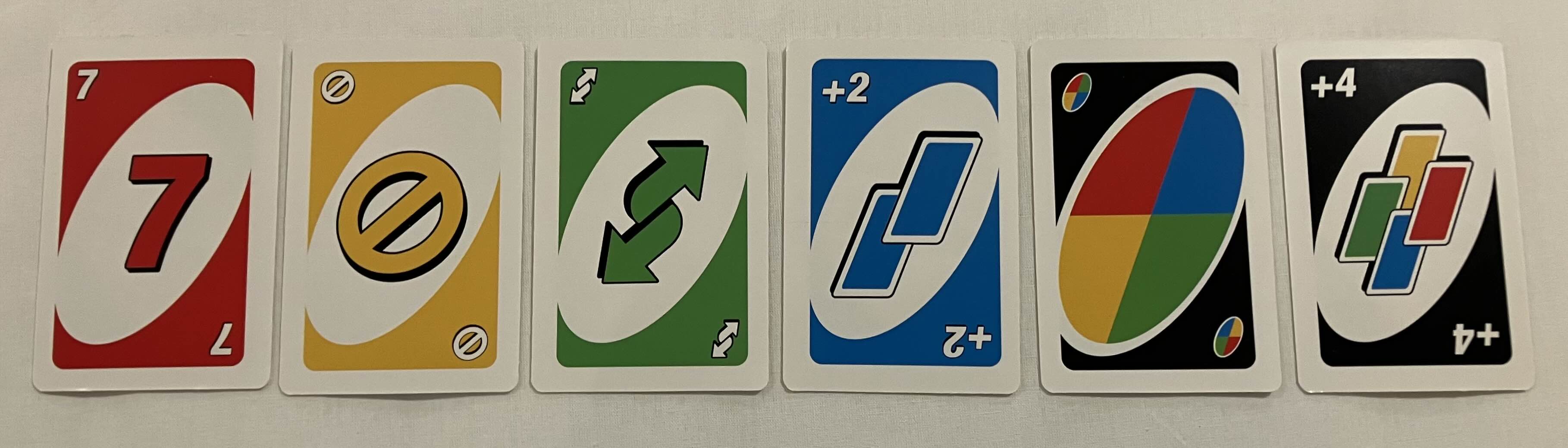}
\caption{From left to right: \crd{$7^R$}, \crd{$S^Y$}, \crd{$R^G$}, \crd{$+^B$}, \crd{$W$}, and \crd{$D$}} \label{fig1}
\end{figure}

A color card may have a single-digit number between 0 and 9, a skip sign, a reverse sign, or a +2 sign written on the front side. All 13 distinct characters can be combined with all four colors, resulting in the total of 52 different types. A black card may have nothing written on the front side (called a \textit{wild card}) or a +4 sign written on the front side (called a \textit{+4 card}). See Fig. \ref{fig1} for examples of UNO cards.

We use \crd{$a^b$} to denote a color card, where $a$ is either a single-digit number, an~`S' for a skip sign, an `R' for a reverse sign, or a `+' for a +2 sign, and $b$ is an~initial of a color (R for red, Y for yellow, G for green, and B for blue). We also use \crd{$W$} and \crd{$D$} to denote a wild card and a +4 card, respectively.

An UNO deck consists of two copies of each type of color card (except for cards with a number 0 with only one copy), and four copies of each type of black card, thus consisting of 108 cards.

Note that our protocol uses additional cards besides the ones in the currently played deck, so several UNO decks have to be prepared beforehand.

\subsection{Pile-Scramble Shuffle}
Given $mk$ cards divided into $m$ piles, each with $k$ cards, a \textit{pile-scramble shuffle for $m$ piles} \cite{scramble} rearranges all piles by a uniformly random permutation unknown to all parties. It can be implemented in real world by putting all cards in each pile into an envelope, and scrambling all envelopes together completely randomly on a table.

The pile-scramble shuffle is denoted by $[p_1|p_2|...|p_m]$, where $p_1,p_2,...,p_m$ are the $m$ piles. For example, a pile-scramble shuffle for two piles:
$$\left[\, \acard{1}\, \acard{2}\, \acard{3}\, \left|\ \acard{4}\, \acard{5}\, \acard{6}\, \right]\, \right.$$
results in either $\acard{1}\, \acard{2}\, \acard{3}\, \acard{4}\, \acard{5}\, \acard{6}$ or $\acard{4}\, \acard{5}\, \acard{6}\, \acard{1}\, \acard{2}\, \acard{3}$, each with probability 1/2.

\subsection{Encoding a Boolean Bit}
Let $\blk$ and $\red$ denote any two different pre-designated types of UNO card (e.g. one may designate a \crd{$1^R$} as $\blk$ and a \crd{$2^G$} as $\red$). A Boolean bit is encoded by a sequence of two cards:
$$\blk\, \red = 0, \quad \red\, \blk = 1.$$

\subsection{Six-Card AND Protocol} \label{6card}
Given two pairs of face-down $\blk$ and $\red$ encoding $x,y \in \{0,1\}$ and two additional cards $\blk\, \red$, a \textit{six-card AND protocol} outputs two pairs of face-down cards encoding $x \wedge y$ and $\bar{x} \wedge y$.
$$\commitment{x}\, \commitment{y}\, \blk\, \red\ \xrightarrow{\text{protocol}}\ \commitment{x \wedge y}\, \commitment{\bar{x} \wedge y}\, \blk\, \red$$

This protocol was developed by Mizuki and Sone \cite{mizuki09}. The procedures are as follows.

\begin{enumerate}
	\item Place the two pairs of cards encoding $x$ and $y$, and two additional cards $\blk\, \red$ encoding 0 as follows.
	$$\commitment{x}\, \commitment{0}\, \commitment{y}$$
	
	\item Rearrange the sequence as follows.
	\vspace{0.3in}
	$$\begin{picture}(40,0)
		\multiput(-15,20)(15,0){6}{\back}
		\put(8,15){\vector(2,-1){25}}
		\put(20,15){\vector(-1,-1){12}}
		\put(35,15){\vector(-1,-1){12}}
		\multiput(-15,-8)(15,0){6}{\back}
	\end{picture}$$
	\vspace{0.03in}
	
	\item Apply the pile-scramble shuffle for two piles.	
	$$\left[\, \back\, \back\, \back\, \left|\ \back\, \back\, \back\, \right]\, \rightarrow\ \back\, \back\, \back\, \back\, \back\, \back\, \right.$$
	
	\item Rearrange the sequence as follows.
	\vspace{0.4in}
	$$\begin{picture}(40,0)
		\multiput(-15,26)(15,0){6}{\back}
		\put(8,21){\vector(1,-1){12}}
		\put(20,21){\vector(1,-1){12}}
		\put(35,21){\vector(-2,-1){25}}
		\multiput(-15,-2)(15,0){6}{\back}
	\end{picture}$$
	
	\item Turn over the two leftmost cards. If they are $\blk\, \red$, the middle and right pairs encode $x \wedge y$ and $\bar{x} \wedge y$, respectively. If they are $\red\, \blk$, the middle and right pairs encode $\bar{x} \wedge y$ and $x \wedge y$, respectively. 
	$$\text{(i)}\ \blk\, \red\, \commitment{x \wedge y}\, \commitment{\bar{x} \wedge y} \, \, \, \, \, \, \text{(ii)}\ \red\, \blk\, \commitment{\bar{x} \wedge y}\, \commitment{x \wedge y}$$
\end{enumerate}

\subsection{Modified Covert Lottery Protocol} \label{covert}
We slightly modify the \textit{covert lottery protocol} developed by Shinoda et al. \cite{lottery}.

In the modified protocol, $m$ face-down arbitrary cards $c_1,c_2,...,c_m$ and $m$ pairs of face-down $\blk$ and $\red$ encoding $x_1,x_2,...,x_m \in \{0,1\}$ are given, where each $x_i$ denotes whether $c_i$ is a valid card that can be selected. If there is at least one valid card, this protocol uniformly selects one of them at random (i.e. if there are $k$ valid cards, each one will have probability $1/k$ to be selected) without revealing any card or the number of valid cards. If there is no valid card, i.e. $c_i=0$ for every $i$, this protocol reports that none is valid. The protocol uses four additional cards $\blk\, \blk\, \red\, \red$.

The only difference from the original protocol is that in the original one, if there is no valid card, the protocol nevertheless uniformly selects one of all $m$ cards at random.

The procedure of this protocol are as follows.

\begin{enumerate}
	\item Place each pair of cards encoding $x_i$ below the card $c_i$, making $m$ piles of three cards.
	$$\overset{\acard{c_1}}{\commitment{x_1}}\ \overset{\acard{c_2}}{\commitment{x_2}}\ \ldots\ \overset{\acard{c_m}}{\commitment{x_m}}$$
	
	\item Apply the pile-scramble shuffle for $m$ piles.
	$$\left[\, \overset{\acard{c_1}}{\commitment{x_1}}\ \right|\, \overset{\acard{c_2}}{\commitment{x_2}}\, \left|\ \ldots\, \left|\ \overset{\acard{c_m}}{\commitment{x_m}}\, \right]\, \rightarrow\ \overset{\back}{\commitment{X_1}}\, \overset{\back}{\commitment{X_2}}\, \ldots\, \overset{\back}{\commitment{X_m}}\, \right.$$
	Let $(X_1,X_2,...,X_m)$ denote the permutation of $(x_1,x_2,...,x_m)$ after the shuffle.
	
	\item Place two additional cards $\red\, \blk$ to encode a bit $t=1$. This pair of cards is called a \textit{token}.
	
	\item For each $i=1,2,...,m$, perform the following steps.
	\begin{enumerate}
		\item Apply the six-card AND protocol in Section \ref{6card} with inputs $X_i$ and $t$ (using two additional cards $\blk\, \red$) to create $X_i \wedge t$ and $\bar{X_i} \wedge t$.
		$$\commitment{X_i}\, \commitment{t}\, \blk\, \red\ \rightarrow\ \commitment{X_i \wedge t}\, \commitment{\bar{X_i} \wedge t}\, \blk\, \red$$
		
		\item Place a pair of cards encoding $y_i = X_i \wedge t$ in the pile previously containing $X_i$. Let the other pair of cards encoding $\bar{X_i} \wedge t$ be the new token $t$ to use in the next iteration.
	\end{enumerate}
	
	\item After $m$ iteration, we now have $m$ piles of three cards, with the bottom pair of each pile encoding $y_1,y_2,...,y_m$, where $y_i = x_i \wedge \bar{x_{i-1}} \wedge \bar{x_{i-2}} \wedge ... \wedge \bar{x_{1}}$ for every $i$.
	$$\overset{\back}{\commitment{y_1}}\ \overset{\back}{\commitment{y_2}}\ \ldots\ \overset{\back}{\commitment{y_m}}$$
	
	\item Apply the pile-scramble shuffle for $m$ piles.
	$$\left[\, \overset{\back}{\commitment{y_1}}\ \right|\, \overset{\back}{\commitment{y_2}}\, \left|\ \ldots\, \left|\ \overset{\back}{\commitment{y_m}}\, \right]\, \rightarrow\ \overset{\back}{\commitment{Y_1}}\, \overset{\back}{\commitment{Y_2}}\, \ldots\, \overset{\back}{\commitment{Y_m}}\, \right.$$
	Let $(Y_1,Y_2,...,Y_m)$ denote the permutation of $(y_1,y_2,...,y_m)$ after the shuffle.
	
	\item Turn over all cards encoding $Y_1,Y_2,...,Y_m$. There can be at most one pair of cards encoding 1. If exactly one pair encodes 1, select the card above that pair; if every pair encodes 0, report that no card is valid.
\end{enumerate}

This protocol uses $m+2$ shuffles, one in the AND protocol per each iteration plus two in the main protocol.

Note that in the original protocol of Shinoda et al. \cite{lottery}, Step 4 is performed for only $m-1$ iterations. Then, $y_m$ is set to be $t$ after the $(m-1)$-th iteration to ensure that exactly one of $y_1,y_2,...,y_m$ is 1.

\section{Overview}
\subsection{Rules of UNO}
Due to the popularity of the game, there are dozens of variants of UNO, each one with slightly different rules. The following set of rules is one of the simplest and one of the most played variants.

All players are located around a circle. Each player begins with seven cards in hand randomly drawn from the deck. The rest of the deck, called the \textit{unplayed deck}, are placed face-down at the center.

First, turn over the topmost card of the unplayed deck and put it in the \textit{discard pile}, designating it as the most recently played card. Then, randomly pick a player to start the game. Play initially proceeds in the clockwise order around the circle.

In each turn, a player plays a card in their hand, discarding it face-up to the discard pile. The played card has to be a color card with the same number or color as the most recently played card, or can be any black card (\crd{$W$} or \crd{$D$}).

\begin{itemize}
	\item If a card with a number has been played, the turn ends with no additional action.
	\item If a card with a skip sign is played, the next player misses a turn.
	\item If a card with a reverse sign is played, the play order changes direction (from clockwise to counterclockwise, and vice versa).
	\item If a card with a +2 sign is played, the next player has to draw two cards from the unplayed deck and misses a turn.
	\item If a \crd{$W$} is played, the current player can designate one of the four colors to it; the next player has to play a card with that color (or a black card).
	\item If a \crd{$D$} is played, the next player has to draw four cards from the unplayed deck and misses a turn. Also, the current player can designate one of the four colors to it; the next player has to play a card with that color (or a black card).
\end{itemize}

If the player does not have a valid card in hand to play, or does not wish to play a card, they have to draw a card from the unplayed deck and may optionally play that card if it is valid.

The first player who plays all cards in their hand wins the game.

\subsection{Simulating Virtual Players}
First, we will show the overview of how to simulate virtual players. Full details of the \textit{card selection protocol}, our most important protocol, will be later explained in Section \ref{main}.

For each virtual player $P$, all people jointly operate $P$ as follows.

\begin{itemize}
	\item At the start of the game, draw seven cards from the deck and put them face-down in $P$'s hand without looking at them.
	\item When $P$'s previous player plays a +2 (resp. +4) card, draw two (resp. four) cards from the unplayed deck and put them face-down in $P$'s hand without looking at them. Note that $P$'s turn is skipped.
	\item In $P$'s turn, apply the card selection protocol in Section \ref{main} to select a random valid card from $P$'s hand to play, or report that none exists.
	\begin{itemize}
		\item If a color card has been played, no further action is necessary.
		\item If a black card (\crd{$W$} or \crd{$D$}) has been played, designate a color of that card by uniformly selecting one of the four colors: red, yellow, green, and blue, at random (which can easily be simulated by shuffling a pile of any four different cards and randomly picking one of them).
		\item If no card has been played, draw one card from the unplayed deck and put it face-down in $P$'s hand without looking at it. Then, apply the card selection protocol in Section \ref{main} again to play the newly drawn card if it is valid (this time if no card is played, no further action is necessary).
	\end{itemize}
\end{itemize}

\section{Card Selection Protocol} \label{main}
Suppose there are $n-1$ players. Let $P_1$ be a virtual player in consideration, and $P_2,P_3,...,P_{n-1}$ be other (human or virtual) players. We consider the unplayed deck as the hand of a hypothetical $n$-th player; the cards in this pile must not be revealed to anyone.

Let \crd{$\theta_1$}, \crd{$\theta_2$}, ..., \crd{$\theta_n$} denote any $n$ different pre-designated types of UNO card, which are also different from $\blk$ and $\red$.

All people jointly operate $P_1$ as follows.
 
\begin{enumerate}
	\item Arrange all players' hands and the unplayed deck face-down horizontally, where the unplayed deck is regarded as $P_n$'s hand.
	$$\underbrace{\back\, \back\, \back\, \back\, \back}_{P_1}\, \, \underbrace{\back\, \back\, \back}_{P_2}\, \, \underbrace{\back\, \back}_{P_3}\, \, \cdots\, \, \underbrace{\back\, \back\, \back\, \back\, \back\, \back}_{P_n}$$
	
	\item Place a \crd{$\theta_i$} below each of $P_i$'s cards ($1 \leq i \leq n$).
	$$\cardbb{?}{$\theta_1$}\, \cardbb{?}{$\theta_1$}\, \cardbb{?}{$\theta_1$}\, \cardbb{?}{$\theta_1$}\, \cardbb{?}{$\theta_1$}\, \, \, \cardbb{?}{$\theta_2$}\, \cardbb{?}{$\theta_2$}\, \cardbb{?}{$\theta_2$}\, \, \, \cardbb{?}{$\theta_3$}\, \cardbb{?}{$\theta_3$}\, \, \cdots\, \, \cardbb{?}{$\theta_n$}\, \cardbb{?}{$\theta_n$}\, \cardbb{?}{$\theta_n$}\, \cardbb{?}{$\theta_n$}\, \cardbb{?}{$\theta_n$}\, \cardbb{?}{$\theta_n$}$$
	
	\item Turn over all face-up cards.
	$$\backbb\, \backbb\, \backbb\, \backbb\, \backbb\, \, \, \backbb\, \backbb\, \backbb\, \, \, \backbb\, \backbb\, \, \cdots\, \, \backbb\, \backbb\, \backbb\, \backbb\, \backbb\, \backbb$$
	
	\item Apply the pile-scramble shuffle to the matrix, with each column as each pile.
	$$\left[\begin{tabular}{c|c|c|c|c|c|c|c|c}
		\ \backk \ \ & \ \backk \ \ & \ \backk \ \ & \ \backk \ \ & \ \backk \ \ & \ \backk \ \ & \ \backk \ \ & \ \ \ & \ \backk \ \ \\
		\ \backk \ \ & \ \backk \ \ & \ \backk \ \ & \ \backk \ \ & \ \backk \ \ & \ \backk \ \ & \ \backk \ \ & \ $\cdots$ \ \ & \ \backk \ \
	\end{tabular}\right]$$
    	
	\item Turn over all cards in the first row.
	$$\overset{\crdd{$4^G$}}{\back}\, \, \, \, \, \overset{\crdd{$7^R$}}{\back}\, \, \, \, \, \overset{\crd{$6^B$}}{\back}\, \, \, \, \, \overset{\crd{$W$}}{\back}\, \, \, \, \, \overset{\crd{$2^Y$}}{\back}\, \, \, \, \, \overset{\crd{$S^Y$}}{\back}\, \, \, \, \, \overset{\crd{$1^B$}}{\back}\, \, \, \, \, \cdots\, \, \, \, \, \overset{\crd{$+^R$}}{\back}$$
	
	\item For each column, if the card in the first row is valid, place $\red\, \blk$ below that column; otherwise, place $\blk\, \red$ below that column. (In the following example, the most recently played card is \crd{$2^R$}.)
	$$\overset{\crd{$4^G$}}{\vala}\, \, \overset{\crd{$7^R$}}{\valb}\, \, \overset{\crd{$6^B$}}{\vala}\, \, \overset{\crd{$W$}}{\valb}\, \, \overset{\crd{$2^Y$}}{\valb}\, \, \overset{\crd{$S^Y$}}{\vala}\, \, \overset{\crd{$1^B$}}{\vala}\, \, \cdots\, \, \overset{\crd{$+^R$}}{\valb}$$

	\item Turn over all face-up cards.
	$$\overset{\back}{\backc}\, \, \overset{\back}{\backc}\, \, \overset{\back}{\backc}\, \, \overset{\back}{\backc}\, \, \overset{\back}{\backc}\, \, \overset{\back}{\backc}\, \, \overset{\back}{\backc}\, \, \cdots\, \, \overset{\back}{\backc}$$
	
	\item Apply the pile-scramble shuffle to the matrix, with each column as each pile.
	$$\left[\ \overset{\back}{\backc}\, \left|\ \overset{\back}{\backc}\, \left|\ \overset{\back}{\backc}\, \left|\ \overset{\back}{\backc}\, \left|\ \overset{\back}{\backc}\ \right|\, \overset{\back}{\backc}\ \right|\, \overset{\back}{\backc}\ \right|\, \cdots\ \right|\, \overset{\back}{\backc}\ \right]$$
	
	\item Turn over all cards in the second row.
	$$\overset{\back}{\valf}\, \, \overset{\back}{\valc}\, \, \overset{\back}{\valc}\, \, \overset{\back}{\valh}\, \, \overset{\back}{\vale}\, \, \overset{\back}{\valc}\, \, \overset{\back}{\valh}\, \, \cdots\, \, \overset{\back}{\vald}$$
	
	\item Consider only columns containing a \crd{$\theta_1$}. In such a column, the card in the first row is $P_1$'s card, and the pair of cards in the third row indicates whether the card in the first row is valid. Remove these columns from the matrix and use the cards in the first and third rows of the these columns as inputs for the modified covert lottery protocol in Section \ref{covert} to randomly select a valid card, or report that none exists.
	$$\backcc\, \, \backcc\, \, \cdots\, \, \backcc\ \, \xrightarrow{\text{modified covert lottery}} \, \, \, \back\, \, \, \text{or none}$$
	
	\item Put the unselected cards from the modified covert lottery protocol back into $P_1$'s hand.
	
	\item For the rest of the matrix, put each card above a $\crd{$\theta_i$}$ back into $P_i$'s hand.
	$$\overset{\back}{\valf}\, \, \overset{\back}{\valh}\, \, \overset{\back}{\vale}\, \, \overset{\back}{\valh}\, \, \cdots\, \, \overset{\back}{\vald}$$
\end{enumerate}

Let $k$ and $k_1$ be the number of remaining cards in the game and the number of cards in $P_1$'s hand, respectively. This protocol uses $3k+4$ additional cards and uses $k_1+4$ shuffles ($k_1+2$ in the modified covert lottery protocol and two in the main protocol).

Note that Step 5 of this protocol reveals the remaining cards in the game, as the sequence in Step 1 excludes the discard pile. Since the discarded cards are publicly known, the remaining cards are also public, so this does not cause a security problem. However, typical human players may not remember which cards remain unless they possess perfect memory. To enhance the enjoyment of the game, it may be preferable to consider the discard pile as $P_{n+1}$'s hand in Step~1 and put a $\theta_{n+1}$ card below each discarded card in Step 2, thereby concealing the remaining cards from all players as well.

\subsection{Proof of Correctness and Security}
First, consider the covert lottery protocol. Correctness and security of the original protocol has been proved in \cite{lottery}. Our modified protocol is almost the same as the original one, except that it reports that no card is valid when that is the case. In such cases, the virtual player has to draw a card from the unplayed deck, so the information that no valid card exists is public. Therefore, the modified covert lottery protocol is correct and secure.

Next, we will prove the correctness of the card selection protocol. Due to the way we place cards in Steps 2 and 6, for each card $c$ in the first row, the card \crd{$\theta_i$} in the second row indicates that $c$ is $P_i$'s card, and the pair of cards in the third row indicates whether $c$ is valid ($\blk\, \red$ for yes and $\red\, \blk$ for no). Therefore, in Step 10 we correctly select all cards in $P_1$'s hand and the cards indicating whether they are valid as inputs for the modified covert lottery protocol. Moreover, in Step 12 each card above a \crd{$\theta_i$} is $P_i$'s card, so we correctly return it to $P_i$. Hence, our protocol is correct.

The card selection protocol follows the computational model of card-based cryptography \cite{formal}, where the security is proved based on information theory. To prove the security of this protocol, it is sufficient to prove that the cards that are turned face-up during the protocol does not reveal any information beyond the public information.

\begin{itemize}
	\item In Step 5, all cards in the first row, i.e. all remaining cards in the game, are revealed. Due to the pile-scramble shuffle, the order of the $k$ columns are uniformly distributed among all $k!$ permuation. Therefore, the only information revealed is the set of these cards, which is public information because it is equal to a full UNO deck minus the public discard pile.
	\item In Step 9, all cards in the second row are revealed. Due to the pile-scramble shuffle, the order of the $k$ columns are uniformly distributed among all $k!$ permuation. Therefore, the only information revealed is the number of \crd{$\theta_i$}s for each $i=1,2,...,n$, which is public information.
\end{itemize}

Hence, the protocol is correct and secure.

\section{Application to Other Games}
The card selection protocol in Section \ref{main} is a generic protocol. It can be used to simulate virtual players in other turn-based card or tile games where each player has to select a valid card to play in each turn. We hereby show three examples of Sevens, Hearts, and Dominoes.

\subsection{Sevens}
Sevens is a card game played with a standard 52-card deck. The whole deck is divided (roughly) equally to all players. A player who has a \crd{$7^\diamondsuit$} starts the game by playing the \crd{$7^\diamondsuit$}. Then, the next player has to play a \crd{$6^\diamondsuit$}, a \crd{$8^\diamondsuit$}, or a \crd{7} of any other suit. In general, in each turn a player may either

\begin{enumerate}
	\item play a \crd{7} of any unplayed suit, or
	\item play a card of a played suit, maintaining that the played cards in that suit still form a set of consecutive ranks. For example, if $\crd{$5^\diamondsuit$}\, \crd{$6^\diamondsuit$}\, \crd{$7^\diamondsuit$}\, \crd{$8^\diamondsuit$}$ have already been played, a player may only play a \crd{$4^\diamondsuit$} or a \crd{$9^\diamondsuit$}.
\end{enumerate}

If a player does not have a valid card to play, they have to pass their turn. The first player who plays all cards in their hand wins the game.

Our card selection protocol works in the same way for Sevens, except that there is no unplayed deck. In each virtual player's turn, all people jointly apply the card selection protocol to select a random valid card in its hand to play, or pass the turn if none exists.

\subsection{Hearts}
Hearts is a four-player card game also played with a standard 52-card deck. Every player begins each round with 13 cards in hand.

A round of Hearts consists of 13 \textit{tricks}. In each trick, every player plays a card in the clockwise order. The first player can play any card. The subsequent players have to play a card having the same suit as the first card if possible; if they do not have such a card, they can play any card. A player who plays the highest-ranked card with the same suit as the first card \textit{wins} the trick, and receives points equal to the points of all four played cards combined. That player also starts the next trick.

Each heart card has one point, and a \crd{$\text{Q}^\spadesuit$} has 13 points. After the end of each round, the points are tallied and the next round begins. The first player who reaches 100 points loses the game.

Our card selection protocol works in the same way for Hearts (without the unplayed deck), except that the original covert lottery protocol \cite{lottery} is used instead of the modified one. In each virtual player's turn, all people jointly apply the card selection protocol to select a random valid card in its hand, or randomly select any card if no card is valid.

\subsection{Dominoes}
Dominoes is a deck of rectangular tiles, each with two numbers between 0~and~6 on its face. The backs of the dominoes are indistinguishable, just like playing cards. Tiles with the same number (such as 6 and 6) are called doublets. Although there are several games of dominoes, \textit{Muggins} (also known as \textit{All Fives}) is one of the most commonly played games of dominoes.

At the beginning of the game, the deck of dominoes is completely shuffled face down and each player starts with five tiles drawn from the deck. The player with the highest doublet plays first, and turns proceed in a clockwise direction. The first player plays any domino, and each player plays a matching tile on one of the endpoints. For example, if the first player plays a tile of $(5,6)$, the second player can play the tile of $(5,x)$ or $(6,y)$. A player who cannot play must repeatedly draw a tile domino from the deck until they can play. A player scores points when playing a tile whose sum of all open endpoints is a multiple of five. The game ends when a player has no tiles left or when all players have passed. The player with the highest score wins the game.

Our card selection protocol works the same way for the game. On each virtual player's turn, all people jointly apply the protocol to select a random valid tile in its hand to play. If there is no valid tile, a tile is drawn from the deck and the protocol is run again.

Note that the selection of the first player can also be solved by our card selection protocol as follows. For simplicity, assume that the number of virtual players is one. First, each real player announces the largest doublet in their hand. For example, suppose that the largest doublet among the real players is $(4,4)$. Then, by specifying that the valid tiles are doublets of $(5,5)$ or $(6,6)$, our card selection protocol tells whether the virtual player has the largest doublet or not. Even if there are two or more virtual players, we can determine the player who has the largest doublet by applying our card selection protocol multiple times.

\section{Future Work}
We developed a card-based protocol that can simulate virtual players to play UNO, as well as other turn-based card games. An interesting future work is to design a protocol to simulate virtual players that can randomly select a card following a non-uniform distribution depending on the available cards, making them more fun and more difficult to play against. Other possible future work includes developing a simuation protocol for card games where a player can simultaneously play multiple cards in each turn, such as \textit{Dominion}.

\end{document}